\documentclass[conference]{IEEEtran}
\usepackage[T1]{fontenc}
\usepackage{blindtext, graphicx,amsmath}
\usepackage[usenames, dvipsnames]{color}
\usepackage{graphicx,graphics}
\usepackage{float}
\usepackage{epsfig}
\usepackage{subfigure}
\usepackage{multirow}

\usepackage{cite}

\ifCLASSINFOpdf
\else
\fi
\begin{document}
%
\title{Indoor and Outdoor Penetration Loss Measurements at $73$ and $81$ GHz}

	

\author{\IEEEauthorblockN{Mahfuza Khatun$^{\dag}$, Changyu Guo$^{\dag}$,  David Matolak$^{\ddag}$, Hani Mehrpouyan$^{\dag}$}\\
	
	\IEEEauthorblockA{$^{\dag}$Department of Electrical and Computer Engineering, Boise State University \\
		$^\ddag$Department of Electrical Engineering, University of South Carolina  \\
		Email: \{mahfuzakhatun, changyuguo,  hanimehrpouyan\}@boisestate.edu, matolak@cec.sc.edu	  
	}
}

\maketitle
\thispagestyle{empty}
\pagestyle{empty}
\begin{abstract}
 In this paper, we present millimeter-wave (mmWave) penetration loss measurements and analysis at E-bands\textemdash $73$ GHz and $81$ GHz. Penetration loss was measured for common building materials such as clear glass, metal, tinted glass, wood, and drywall on the campus of Boise State University in the city of Boise. A horn antenna with a gain of $24$ dBi was used at the transmitter and receiver at both bands, and both antennas were boresight-aligned with respect to the test material. A total of twelve locations were selected to test five materials. We tested two indoor materials (clear glass and wood) in at least two locations to determine the effect of penetration loss of materials in similar compositions. The average penetration loss and standard deviation were estimated for these indoor materials. We measured an average penetration loss of $2$ to $9$ dB for wood and glass, respectively. Furthermore, we measured the penetration loss of common indoor and outdoor building materials. We studied that outdoor materials had larger penetration losses, e.g., we obtained a penetration loss of $22.69$ dB for outdoor metal, where this value dropped to $16.04$ dB for indoor metal at $73$ GHz. Similar results were also obtained for the $81$ GHz channel, where the largest penetration loss was measured to be $26.5$ dB through a tinted glass door in an outdoor setting.
\end{abstract}
\IEEEpeerreviewmaketitle


\section{Introduction}

Due to the rapid growth of new technologies, wireless networks are facing new challenges. The future wireless networks are expected to support high throughput links with a progressively distinct set of applications. However, the available spectrum in the sub-$6$ GHz spectrum will not be enough to support these new applications. To overcome these challenges, the millimeter-wave spectrum has been topic of significant research due to the enormous available spectrum in this band~\cite{rqppaport2013Itwillwork,rangan2014millimeter,rappaport2015wideband,
	rappaport2014millimeter}.

To fully use the potential of the spectrum in the mmWave band and achieve higher data rates and reliable wireless links, a comprehensive knowledge of the channel models in both indoor and outdoor settings are needed. This includes propagation characteristics, such as path-loss, penetration loss, delay spread, and angular spread. In this paper, we focus our attention on developing outdoor and indoor setups for penetration measurement of building materials at two mmWave bands\textemdash$73$ GHz and $81$ GHz. Knowledge of penetration measurements of common building materials at mmWave frequency bands is needed to ensure reliable indoor-to-outdoor and outdoor-to-indoor wireless links~\cite{rappaport2014millimeter}. Moreover, there are shortage of penetration measurements at high frequency bands. These two reasons are the main motivations behind our research work in this paper.





Accordingly, over the past few years, several companies and research groups have been carrying out channel propagation measurements at mmWave frequencies, e.g., $28$, $60$, $ 73$, and $81$ GHz in line-of-sight, none-line-of-sight, point-to-point, indoor and outdoor scenarios~\cite{rappaport2015wideband, samimi20153, mahfuza2017, mahfuza2018,mahfuza_vtc}. For instance, wideband channel measurement campaigns were performed in the $81$-$86$ GHz bands for both a street canyon and root-to-street scenarios~\cite{kyro2010long, kyro2012experimental}.A high gain directional antenna with proper structure \cite{shad2019} are used to mollify propagation losses at high frequencies. 
However, there have been relatively few published studies for penetration loss at mmWave frequencies as mentioned earlier. Rappaport \textit{et al.} showed measurements for reflection and penetration loss of common building materials in dense urban environments at $28$ GHz\cite{zhao201328,rqppaport2013Itwillwork}. Another study on the propagation path loss in a building at $60$ GHz was conducted in~\cite{1296643}. The results in~\cite{1296643} showed that at $60$ GHz due to large penetration and pathloss the signal can be effectively confined to a single room. Additionally, Rappaport \textit{et al.} observed that penetration loss at $73$ GHz does not necessarily increase and decrease based on the antenna polarization configuration~\cite{ryan2017indoor}. In~\cite{ryan2017indoor} a measurement campaign was also performed in a typical indoor office environment at $73$ GHz. Unlike prior work in the literature, we present for the first time penetration losses for the $73$ and $81$ GHz spectrum for both indoor and outdoor building materials. 

In this paper, we present the results of a comprehensive penetration loss measurement campaign for both indoor-to-indoor and indoor-to-outdoor at the E-band. The propagation measurement campaigns were performed in and around the buildings on the campus of Boise State University (BSU). Our measurement results show that the outdoor building materials result in a higher penetration loss compared to that of the indoor for both $73$ and $81$ GHz.

This paper is structured as follows: Section~\ref{sec:equipment} describes the experimental equipment setup for penetration loss measurements. In Section~\ref{sec:environment}, we describe the building environments in which the measurement campaign was carried out at. Section~\ref{sec:methods} briefly summarizes the measurement procedure used throughout this paper. In Section~\ref{sec:result}, we provide the results collected from the penetration measurement campaigns at the two frequency bands. Finally, Section~\ref{sec:conclusion} concludes this paper.

\begin{figure}[t]
	\begin{center} 		
		\includegraphics[width=0.9\linewidth]{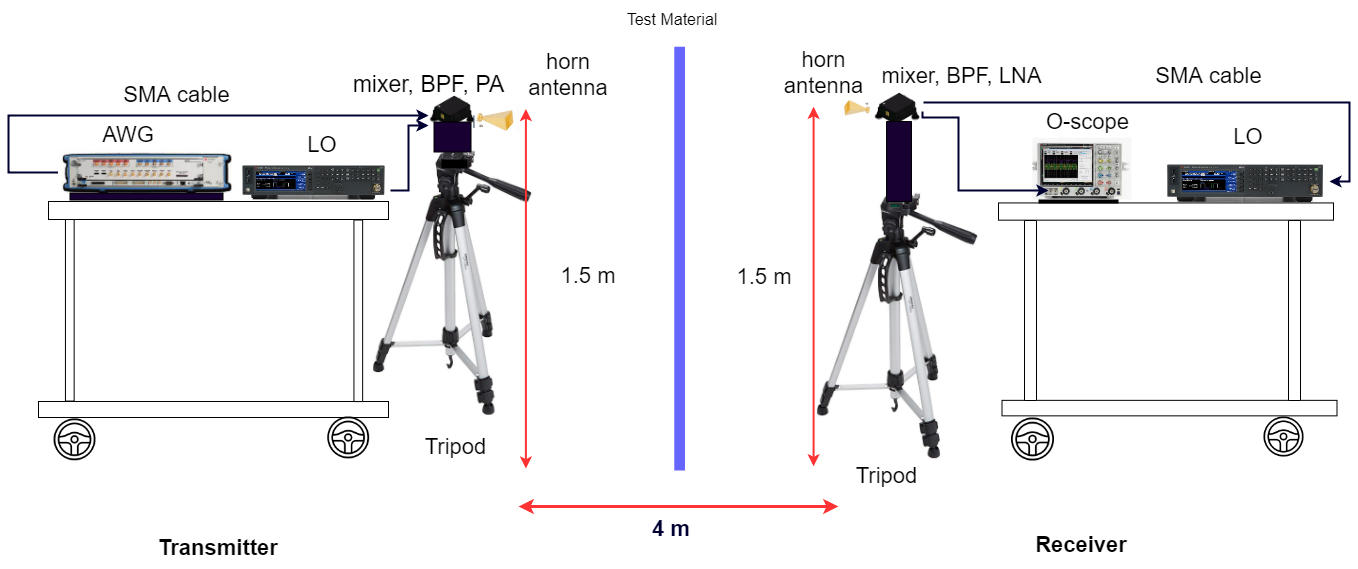}
	\end{center}
	\caption{ Test setup for penetration measurement at $73$ GHz}
	\label{Fig1}
\end{figure}

\section{Measurement Setup}
\label{sec:equipment}
For the $73$ GHz penetration measurement campaign, we used a wideband arbitrary waveform generator (AWG) as the transmitter (TX), and an oscilloscope as the receiver (RX). The waveform generator and scope were connected and synchronized using a $10$ MHz signal. At the transmitter, the signal was digitally modulated at $4$ GHz with binary phase shift keying (BPSK) modulation. Following this, the intermediate frequency (IF) signal from the AWG was mixed with a local oscillator to reach $73$ GHz. Subsequently, we used a band pass filter (BPF) to remove the unwanted signals. Later, a power amplifier (PA) with a gain of $20$ dB was placed before the TX antenna. A horn antenna was used with a gain of $24$ dBi and a $3$ dB beamwidth of $11^{\circ}$ and $7^{\circ}$ in the azimuth and the elevation angles, respectively. The transmit power was measured at $2$ dBm. At the receiver, a scope, a downconverter, a local-oscillator, a low noise amplifier, and a band pass filter were employed for receiving the original signal at $4$ GHz. An identical horn antenna was used at the RX. The received signal is then fed to the scope, where the received signal strength is estimated. At each location, five measurements were made for averaging purposes. Fig.~\ref{Fig1} shows the measurement test setup at $73$ GHz, where both the TX and the RX were placed atop instrument carts.

For the $81$ GHz channel measurement campaign, we used the same hardware settings, which was employed for measuring the $73$ GHz penetration loss. In addition to the above mentioned equipments, we used a different bandpass filter with the appropriate passband. The TX and RX antennas were identical to before. The transmit power was set and measured to be $1$ dBm. All antennas were placed in vertical-to-vertical polarization for both $73$ GHz and $81$ GHz.

\begin{figure}[t]
	\begin{center} 		
		\includegraphics[width=0.9\linewidth]{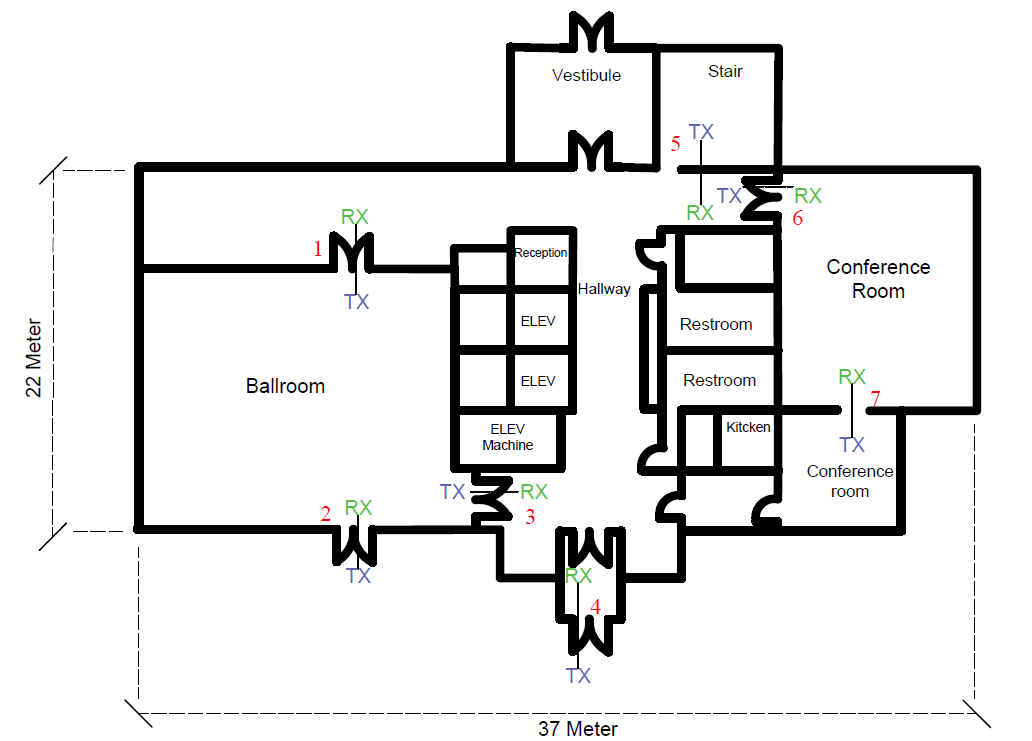}
	\end{center}
	\caption{Floor plan of the 1st floor of Alumni and Friends Center, with TX and RX locations. Note that the doors were closed during measurements}
	\label{Fig2_map}
\end{figure}

\begin{figure}[t]
	\begin{center}
		\subfigure [ ]
		{ \includegraphics[width=0.47\linewidth,height=5cm]{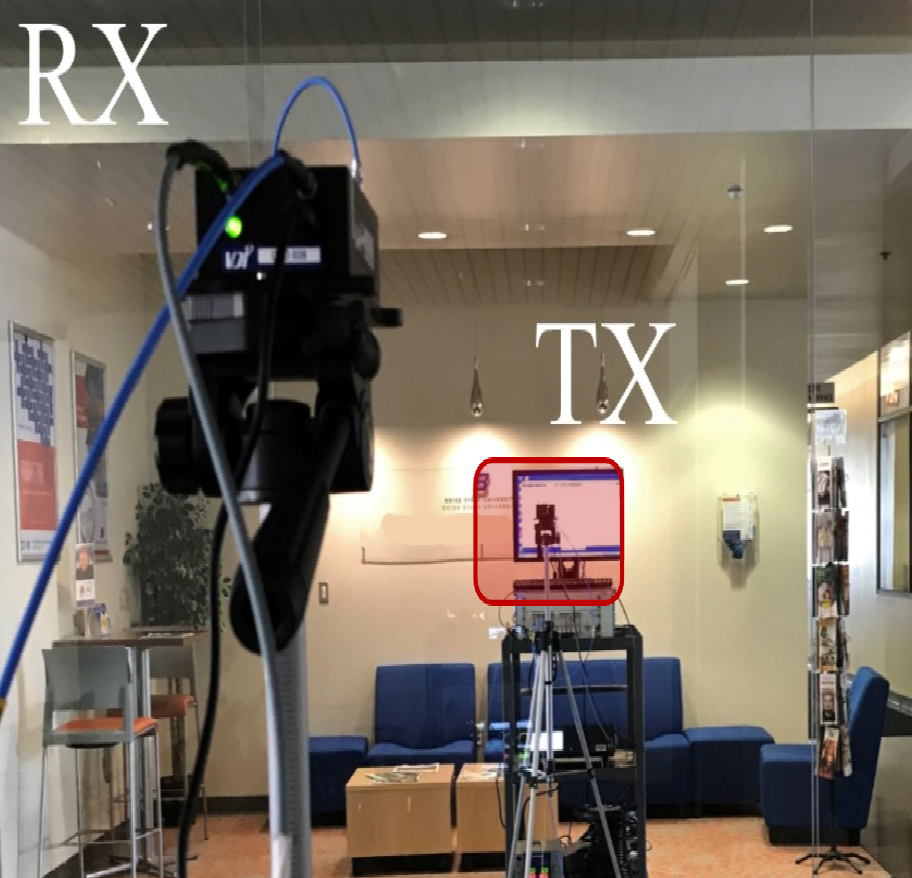}\label{fig:clearglassindoor1}}
		\subfigure []
		{\includegraphics[width=0.47\linewidth,height=5cm]{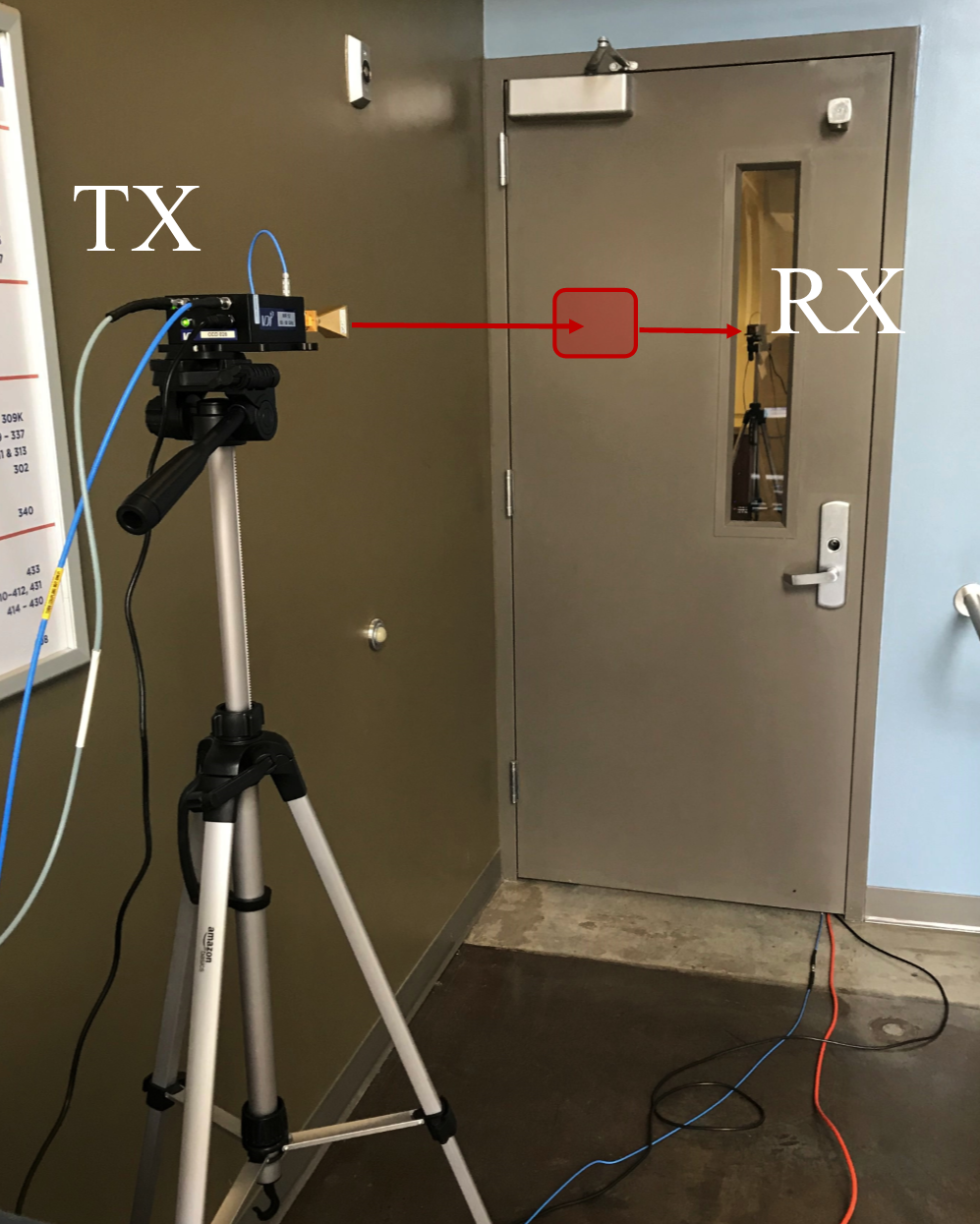}\label{fig:metaldoor}}
	\end{center}
	\caption{Penetration measurement setup in an indoor environment at Norco building to test (a) clear glass wall (b) metal door. Both TX and RX antennas were bore-sighted on a material indicated by a red box}
	\label{Fig_4}
	
\end{figure}	

\section{Environment Descriptions}
\label{sec:environment}
During the spring of 2019, the penetration measurements were performed using the setup mentioned in the prior section. Measurements for common building materials were conducted at three buildings on campus at BSU in the city of Boise. These are: 1) the Alumni and Friends Center (AFC), 2) Micron Engineering (MEC), and 3) Norco Building (NB). The Alumni and Friends Center is a four-floor building with tinted glass walls. Our measurements here pertain to the first-floor, which resembles a typical office environment, with wood doors, tinted glass doors, and tiled/carpeted floors. There are several conference and ball rooms on both sides of the hallway. The door is made of solid core fire rated wood with a thickness of $4.5$ cm. The overall layout is shown in Fig.~\ref{Fig2_map}.

Moreover, we performed another set of measurements on the first-floor of the Norco Building, which exhibits a similar setting to an office environment with metallic doors, clear glass walls and drywall. Measurements were made at three locations of two materials, i.e., clear glass and metal doors. Fig.~\ref{Fig_4} shows two pictures of the setup to test the penetration loss for clear glass and metal at the Norco Building.

\begin{figure}[t]
 	\begin{center}
 		\subfigure [ ]
 		{ \includegraphics[width=0.47\linewidth,height=5cm]{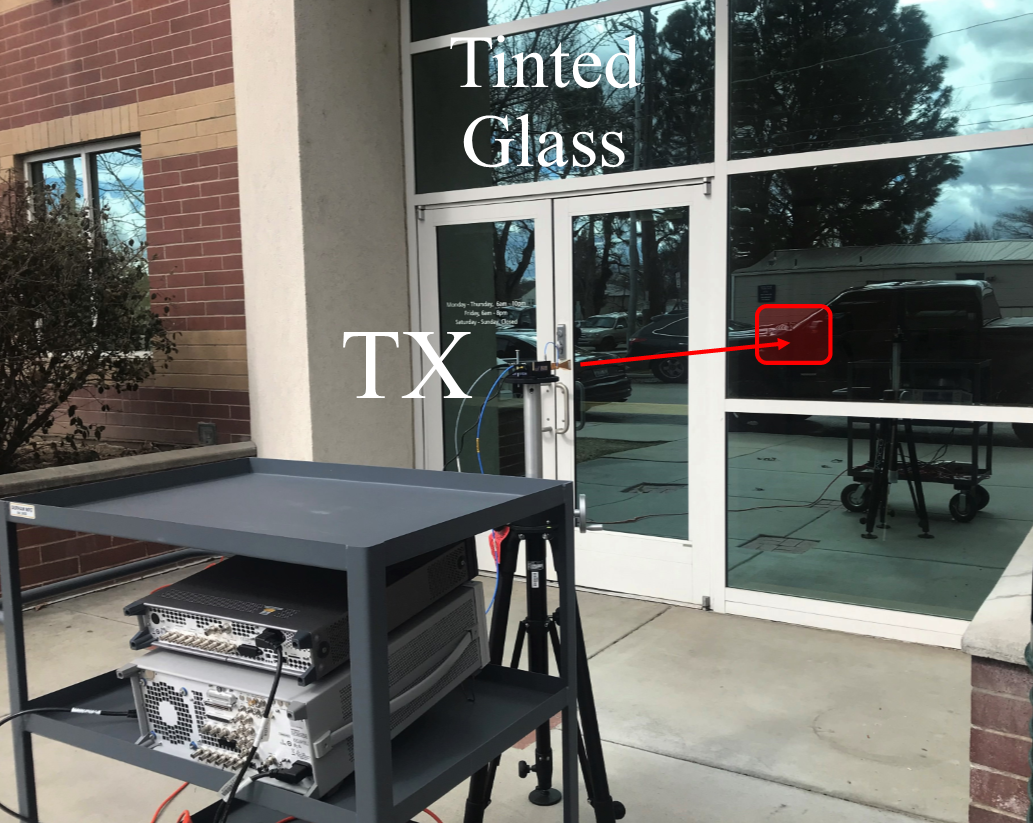}\label{fig:firedoorpenetrationinside}}
 		\subfigure []
 		{\includegraphics[width=0.47\linewidth,height=5cm]{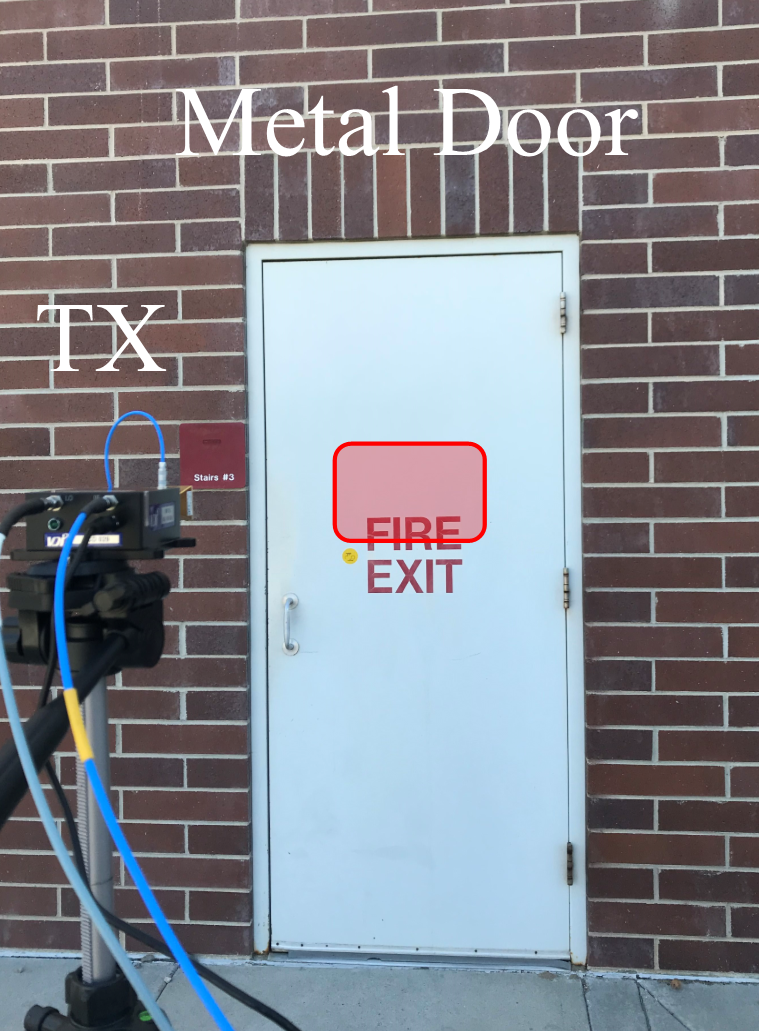}\label{fig:metal}}
 	\end{center}
 	\caption{Penetration measurement setup in an outdoor environment at MEC to test (a) tinted glass (b) metal door. Both TX and RX antennas were bore-sighted on a material indicated by a red box}
 	\label{Fig_5}
 	
 \end{figure}

Our last measurement campaign was performed on the first-floor of the Micron Engineering building. The floor has a long hallway with several laboratories and seminar rooms. The walls are made of sheetrock over metal studs, the ceiling tiles are made of a fiberboard material, and the ground is comprised of concrete. Figs.~\ref{fig:firedoorpenetrationinside} and~\ref{fig:metal} show the penetration measurement setup for a tinted glass and metal, respectively, in an outdoor environment. As shown in Fig.~\ref{fig:firedoorpenetrationinside}, the transmitter was placed outside the building and the RX was positioned inside the building, and both the TX and the RX antennas were pointed on a tinted glass wall.

\begin{table*}[ht]
	
	\renewcommand{\arraystretch}{1.2}

	\centering
	\caption{Averages penetration loss with standard variation at a distance, $4$ m. $\sigma$ is the standard deviation of the penetration loss over all locations of a same type of material}
	\label{tab:my-table}
	\begin{tabular}{|c|c|c|c|c|c|c|}
		\hline
		\multicolumn{7}{|c|}{Average penetration loss of several indoor building materials}                                                                                                                                \\ \hline
		Frequency                     & Material                     & Thickness,cm & Locations & No of location & \begin{tabular}[c]{@{}l@{}} Average penetration loss, dB\end{tabular} & Standard deviation $\sigma$,dB               \\ \hline
		\multirow{4}{*}{73 GHz} & \multirow{2}{*}{clear glass} &       $0.6$    & AFC      & 1              & \multirow{2}{*}{2.66}                               & \multirow{2}{*}{0.02} \\ \cline{3-5}
		&                              &        $1.0$   & NB       & 2              &                                                     &                         \\ \cline{2-7} 
		& \multirow{2}{*}{wood door}   &       $4.6$    & AFC      & 3              & \multirow{2}{*}{7.57}                               & \multirow{2}{*}{2.1}    \\ \cline{3-5}
		&                              &   $4.2$        & MEC      & 1              &                                                     &                         \\ \hline
		\multirow{4}{*}{81 GHz} & \multirow{2}{*}{clear glass} &     $0.6$        & AFC      & 1              & \multirow{2}{*}{}                                   & \multirow{2}{*}{}       \\ \cline{3-5}
		&                              &         $1.0$   & NB       & 2               &   3.18                                                  &   0.56                      \\ \cline{2-7} 
		
		& \multirow{2}{*}{wood door}            &      $4.6$     &  AFC       &          3      & \multirow{2}{*}{9.05}                                   & \multirow{2}{*}{2.4}       \\ \cline{3-5}
		&                              &       $4.2$     &      MEC    &          1      &                                                     &                         \\ \hline
	\end{tabular}

\end{table*}


\section{Measurement Methodology}
\label{sec:methods}
For both $73$ and $81$ GHz penetration loss measurement, the TX and RX were first separated by a certain separation distance. The distance, $4$ m was selected in free-space to provide a far-field reference path loss at the locations of measurement campaigns, as explained in the prior section. Then, the TX and RX were moved to be on each side of a test material while still keeping same separation distance. Both TX and RX were placed $2$ m away from a test material. The height of both TX and RX antennas was set at $1.5$ m from the ground level. At each location, we approximately aligned the TX and RX antenna in boresights. We captured five power delay profiles (PDPs) at each TX-RX location, and the setup was kept fixed for each set of five repeated PDPs.

To determine the received power $P_{r}(d)$ at a distance $d$, we recorded the power of the first-arriving multipath component (MPC) from the observed power delay profiles~\cite{ryan2017indoor,durgin1998basic,newhall1997antenna }, because the first-arriving path was supposed to be a penetration path. We paid no attention to later-arriving paths as they considered to come from reflections caused by nearby obstacles in the environment~\cite{joshi2005near}. 

Penetration loss was computed as the difference between average received power under a test material and the received power in unobstructed free-space with the same TX-RX separation. Along with knowledge of transmit power, the estimated path loss is presented by~\ref{eq1} 
\cite{rappaport1996wireless}.

\begin{align}
\label{eq1}
PL(d)(\text{dB}) = P_t(\text{dBm}) - P_{r}(d)(\text{dBm}) + G_t(\text{dB}) + G_r(\text{dB})
\end{align}

where, $d = 4$ is the TX-RX separation distance. $P_t$ is the transmit power, $G_t$ and  $G_r$ are the gain of transmitter and receiver antennas, respectively. $P_{r}(d)$ is the measured received power at a distance, $d$.
The penetration Loss, $L$ in dB can be obtained using~\ref{eq2}

\begin{align}
\label{eq2}
L(dB) = PL^{unobs.}(d) - PL^{MUT}(d)
\end{align}

where, $PL^{MUT}(d)$ is the path loss for material under test which is denoted by MUT, and $PL^{unobs.}(d)$ is the path loss in the unobstructed free-space distance.


\begin{figure}[t]
	\begin{center}
		
		\includegraphics[width=0.9\linewidth,keepaspectratio]{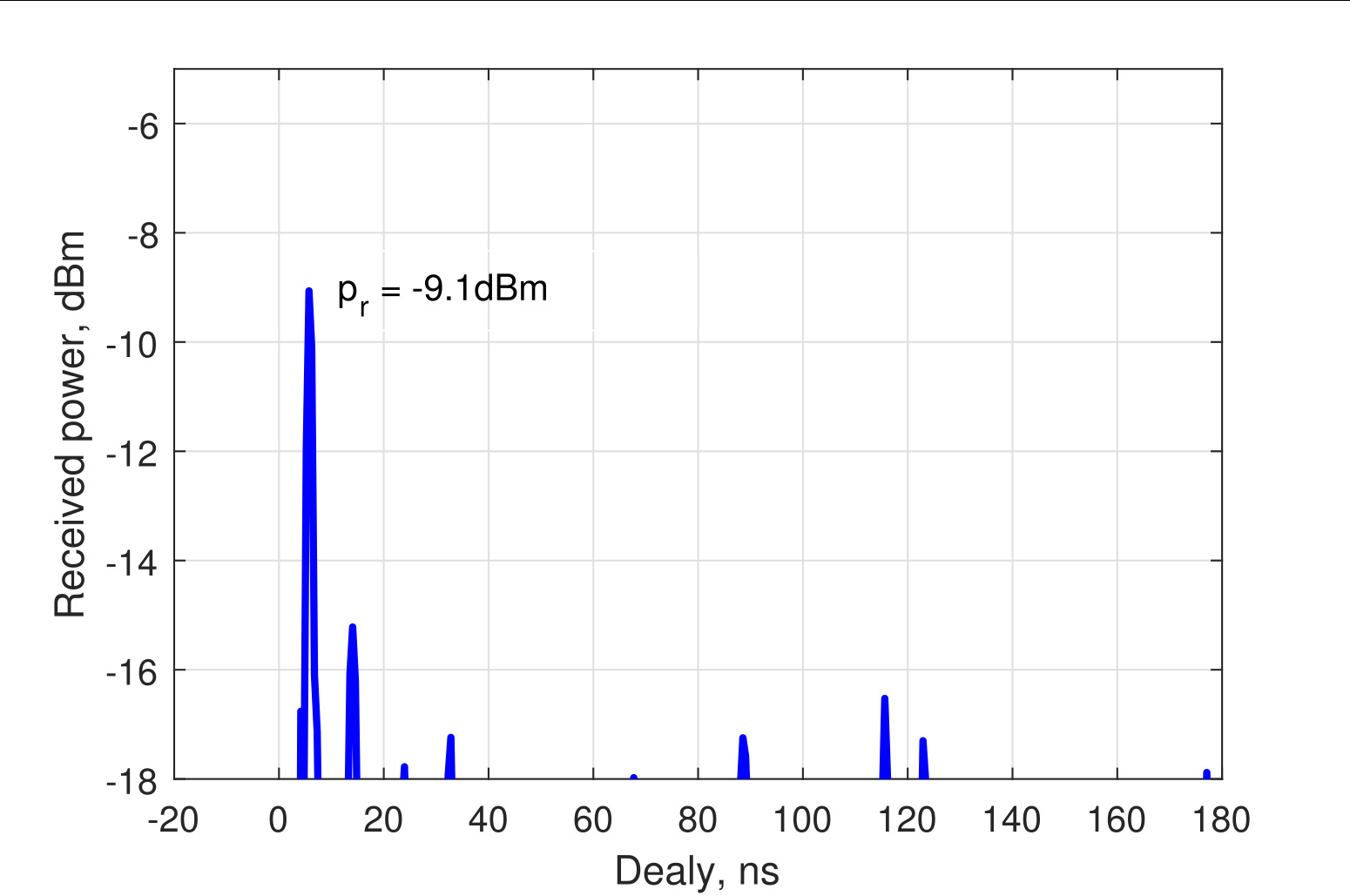}
	\end{center}
	\caption{Power Delay Profile observed in the $4$ m tinted glass penetration test at $73$ GHz at the MEC in an outdoor scenario. $ P_{r} $ is the received power of the first arriving peak}
	\label{fig:pdpclear-glassafc}
\end{figure}

\section{Results}
\label{sec:result}

Table~\ref{tab:my-table} represents a summary of the average penetration losses of common indoor building materials at $73$ GHz and $81$ GHz. The standard deviation of penetration loss for all the locations of each indoor material are also provided in Table~\ref{tab:my-table}. Same types of materials were tested in at least two locations to characterize the effect of penetration loss with the thickness and formation of the material. At both $73$ GHz and $81$ GHz mmWave channels, the penetration loss was obtained larger for the wood doors
when we compared to the clear glass walls. At $73$ GHz, the average penetration loss was measured to be $7.57$ dB in the case of the wooden door. The thickness of the wood doors was measured, ranging in width from $4.2$ to $4.6$ cm, and the doors are constructed using two layers of solid wood cores. The average penetration
loss for the wooden door at $81$ GHz is increased to $9.05$ dB,
 which was around $2$ dB more attenuation than what was observed at $73$ GHz because the penetration depends not only on the thickness and formation of the material but also on the carrier frequency. On the other hand, clear glass walls were made of single layer glass, ranging in width from $0.6$ cm to $1.0$ cm. This relatively lower thickness of the glass can be attributed to the lower average penetration loss. Furthermore, the wooden door locations had a penetration loss with standard deviation of $2.1$ dB at $73$ GHz and $2.4$ dB at $81$ GHz. The penetration loss standard deviation for clear glass door was less than $1$ dB for both the bands because the thickness variation of this material was negligible (within 0-4mm) at different locations.   

Fig.~\ref{fig:pdpclear-glassafc} shows a power delay profile measuring penetration
loss of a tinted glass at the AFC building in an outdoor
environment. Here, the frequency of operation is $73$ GHz It is also observed that, the line-of-sight (LOS) component arrives at $5$ ns delay to the receiver with a received power of $-9.1$ dB. After the first multipath component through the tinted glass, the second component arrives to the receiver with a delay of $15$ ns. The penetration loss was obtained $4.3$ dB with the consideration of the first multipath component. Moreover it is seen that, later arriving MPCs still have sufficient power to be acquired at the RX. This indicates that there are many reflective obstacles in nearby environment and those components might help to propagate well at $73$ GHz.

\begin{table*}[t]
	\centering
	\caption{Comparison of Penetration Losses of Indoor Materials and Outdoor Materials at $73$ GHz and $81$ GHz. }
	\label{tab:my-table-2}
	\begin{tabular}{|c|c|c|c|c|c|c|c|}
		\hline
		\multicolumn{8}{|c|}{ Penetration loss for indoor and outdoor building materials }                                                                                                                                                                                                                                                       \\ \hline
		Frequency                 & Environment                  & Locations & Material     & Thickness, cm & \begin{tabular}[c]{@{}l@{}}Received power\\ free-space, dB\end{tabular} & \begin{tabular}[c]{@{}l@{}}Received power\\ material, dB\end{tabular} & \begin{tabular}[c]{@{}l@{}}Penetration Loss, dB\ \end{tabular} \\ \hline
		\multirow{5}{*}{$73$ GHz} & \multirow{3}{*}{Outdoor} & MEC & Metal   &    $4.7$   &  $ -5.22$
		&          $-27.91$
		&                $22.69 $                                                     \\ \cline{3-8} 
		&                      & AFC & \begin{tabular}[c]{@{}l@{}}Tinted \\Glass \end{tabular}      &   $4.2$      & $-6.0 $                                                                     &     $ -32.0  $                                &  $26.0$            \\ \cline{3-8} 
		&                      & MEC & \begin{tabular}[c]{@{}l@{}}Tinted \\Glass \end{tabular}  &    $1.2$      & $-4.8 $                                                                     &     $ -9.1  $                                                           &  $4.3$                          \\ \cline{2-8} 
		
		& \multirow{2}{*}{Indoor}  & NB & Metal        &   $4.5$      &             $-10.12$                                              &     $-26.2 $                &     $16.04  $     \\ \cline{3-8} 
		&                      & AFC & Wall &    $26$      &     $-7.4$                &          $-11.42$                    &    $ 4.02$
		\\ \hline

		\multirow{5}{*}{$81$ GHz} & \multirow{3}{*}{Outdoor} &  MEC & Metal           &   $4.7$        &                         $-8.2$   &   $-33.0$     &   $24.8$                                                 \\ \cline{3-8} 
		
		&                   &   AFC  &     \begin{tabular}[c]{@{}l@{}}Tinted \\Glass \end{tabular}         &   $4.2$ &   $-8.0$   &     $-34.5$ &       $26.5$    \\ \cline{3-8}

		&                      &   MEC      &       \begin{tabular}[c]{@{}l@{}}Tinted \\Glass \end{tabular}                 & $1.2$ &         $-6.2$         &     $-10.8$                                                              &  $4.58$                                                  \\ \cline{2-8} 
		
		& \multirow{2}{*}{Indoor}    &  NB   &     Metal         &    $4.5$      &      $-13.2$ &  $-28.6$   &       $15.4$                                              \\ \cline{3-8} 
		&                      &  AFC   &      Wall    &       $26$    &     $-8.2$             &    $-12.45$    &     $4.25$                                               \\ \hline
	\end{tabular}
	
\end{table*}

\begin{figure}[t]
	\centering
	\includegraphics[width=0.9\linewidth,keepaspectratio]{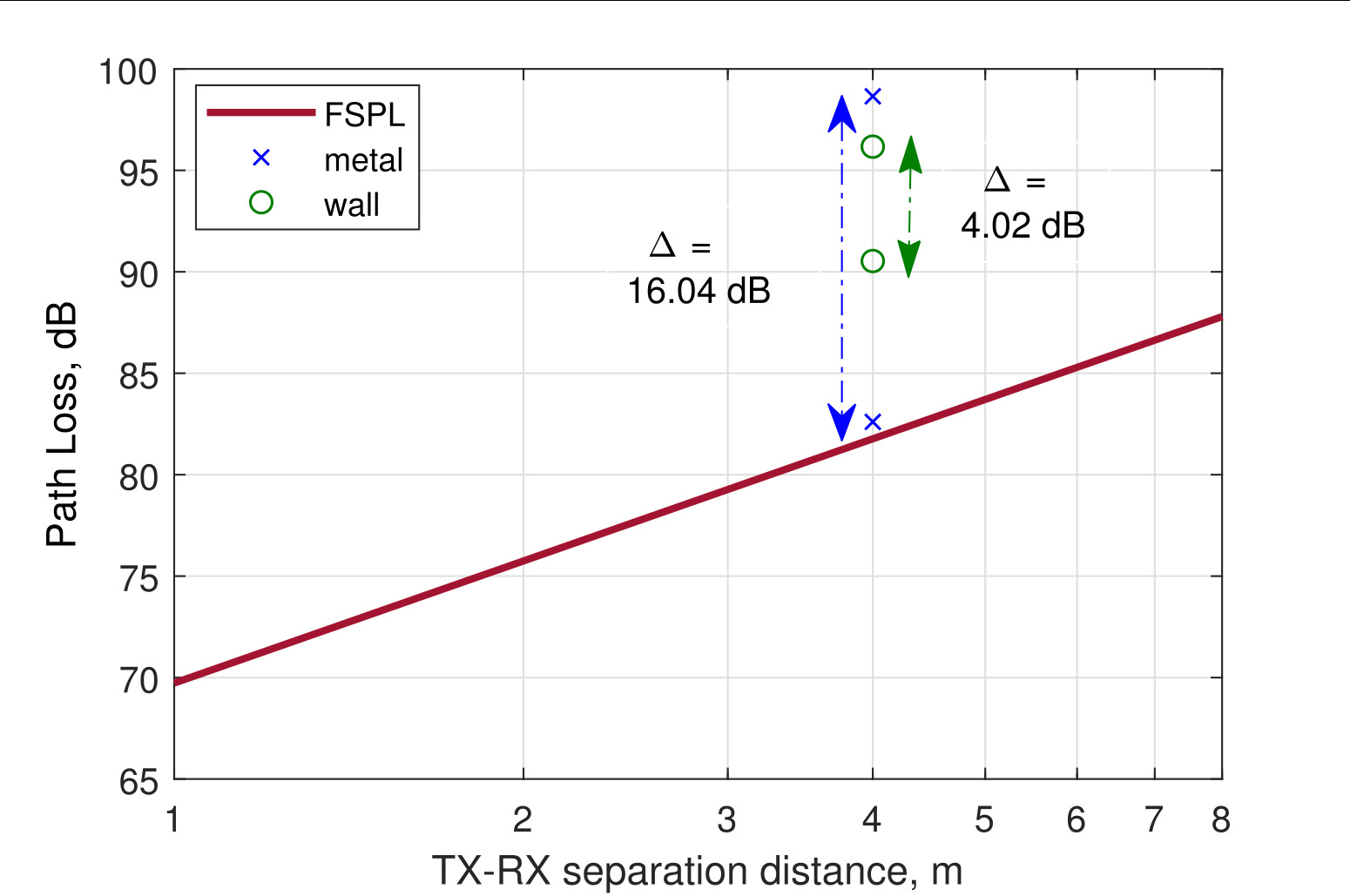}
	\caption{Penetration loss of metal door and drywall at $73$ GHz in indoor scenarios. The blue cross and green circle show the path loss for metal and drywall test, respectively. $\Delta$ is the difference of the $PL^{unobs.}(d)$ and $ PL^{MUT}(d)$ of a test material. The red line presents the free-space path loss at $73$ GHz }
	\label{fig:lossmetalwall73}
\end{figure}


Table~\ref{tab:my-table-2} contains the penetration results of indoor and outdoor building materials at both bands\textemdash$73$ GHz and $81$ GHz\textemdash at different locations. In both frequencies, the outdoor tinted glass door at AFC had the highest penetration loss. The door was composed of two layers of solid tinted glass with a thickness of $4.2$ cm, which resulted the reduced the received signal strength at the receiver. Fig.~\ref{fig:lossmetalwall73} shows the penetration loss measurement results of indoor materials (metal and drywall) at $73$ GHz, where $\Delta$ is the loss for a material at one location. As depicted in Fig.~\ref{fig:lossmetalwall73}, metal shows a higher attenuation of $16.04$ dB at $4.5$ cm thickness for the $73$ GHz indoor channel. Indoor drywall had moderate attenuation of $4.02$ dB at $26$ cm thickness at this band. Comparing the measured loss for both bands, we can observe that the attenuation for metal in an indoor environment decreased from $73$ to $81$ GHz. This is unknown, why the penetration loss through metal in an indoor environment decreases from $73$ to $81$ GHz, as reported in Table~\ref{tab:my-table-2}.

The outdoor building materials had more attenuation than that of the indoor building materials at each frequency band as presented in Table~\ref{tab:my-table-2}. At $81$ GHz, the outdoor metal had higher attenuation of $24.8$ dB, as compared to indoor metal, which had lower penetration of $15.4$ dB. This is most likely due to the formation of the outdoor building materials, which contained thick and dense layers of materials. It is clear that indoor-to-outdoor penetration will be quite difficult at the high frequency, whereas signals at high frequencies can easily propagate in the indoor environment. This indicates that high frequency reuse will be recommended between indoor and outdoor mmWave networks, which minimizes the interference.

\section{Conclusion}
\label{sec:conclusion}
This paper presents the results of penetration loss measurements of common building materials at $73$ GHz and $81$ GHz. We performed channel measurement campaigns at three buildings on the campus of BSU, where the received signal strength was measured for various indoor and outdoor building materials such as clear glass, tinted glass, wood, metal, and drywall. The
standard deviation examined among measurements of the indoor building
materials at various locations was as large as $2.4$ dB for the
wooden door at $81$GHz channel, and the lowest standard
deviation was obtained to be $0.56$ dB through the clear glass
wall for the same frequency. However, at both frequencies
\textemdash $73$ and $81$ GHz, average penetration loss was examined from $2$ to $ 9$ dB for the indoor materials, which were tested in at least two locations. These results suggest that electromagnetic waves
 can penetrate through glass walls or wooden doors reasonably
 well.
 
Moreover, the outdoor building materials had larger penetration loss at $73$ GHz in compared to the indoor building materials. This is most likely due to the thickness and formation of outdoor materials, which cause more attenuation and higher reflection than the indoor materials. These results also indicate that high frequency reuse is needed between indoor and outdoor networks. Future work includes additional measurements for penetration to consider other type of materials at mmWave frequencies as well as reflection coefficient measurements for those materials.

\section*{Acknowledgment}
This research work was funded by NASA$'$s Aeronautics Research Mission Directorate under the University Leadership Initiative. The authors would like to thank Paul Robertson for outstanding help to verify the composition of the building materials during the measurement campaign.

\bibliographystyle{IEEEtran}
\bibliography{IEEEabrv,ref}

\end{document}